\documentclass[twocolumn,amsmath,amssymb,aps,prb]{revtex4-2}
\usepackage{graphicx}
\usepackage{hyperref}
\usepackage{xcolor}
\usepackage{multirow}
\begin{document}

\preprint{APS/123-QED}

\title{Spin Currents in Rashba Altermagnets: From Equilibrium to Nonlinear Regimes}

\author{$^1$Priyadarshini Kapri}
\email{priya.kapri@gmail.com}

\affiliation{Department of Mathematics and Physics, North Carolina Central University, Durham, NC-27707, USA}

\date{\today}

\begin{abstract}

We investigate equilibrium (background), linear, and nonlinear spin currents in two-dimensional Rashba spin-orbit coupled altermagnet systems, using a modified spin current operator that includes anomalous velocity from non-zero Berry curvature. The background spin current, stemming from spin-orbit coupling and modulated by the altermagnet term ($t_j$), exhibits in-plane polarization, increases linearly with Fermi energy ($\epsilon_F$), and is enhanced by both the altermagnet ($t_j$) and the Rashba parameter ($\lambda$).
 Linear spin current is always transverse with out-of-plane polarization and can be viewed as Spin Hall current, primarily driven by band velocity, with 
$t_j$ enabling a band-induced contribution (previously absent in simple Rashba systems ($t_j=0$)). This highlights altermagnet system  as a promising source of spin Hall current generation. 
For  linear spin Hall current, its band contribution's magnitude increases linearly with $\epsilon_F$, while the magnitude of  anomalous component saturates at higher $\epsilon_F$. Further, the magnitude of spin Hall current is enhanced by $t_j$
but reduced by $\lambda$.
Nonlinear spin currents feature both longitudinal and transverse components with in-plane polarization.  Both the nonlinear longitudinal spin current from band velocity and the nonlinear transverse spin current from anomalous velocity initially decrease with $\epsilon_F$
 before saturating at higher $\epsilon_F$.  Importantly, $t_j$
  reduces these currents while $\lambda$ enhances them. Meanwhile, the nonlinear transverse current from band velocity increases and then saturates with $\epsilon_F$, enhanced by $\lambda$ and showing non-monotonic variation with 
$t_j$. These findings highlight the tunability of spin current behavior through Rashba and altermagnet parameters, offering insights for spintronic applications.

\end{abstract}

\maketitle

\section{Introduction}
A recent breakthrough in magnetism involves the discovery of a class of antiferromagnetic materials known as altermagnets \cite{Mazin,Hayami1,Hayami2,Hayami3}. These materials feature strongly spin-polarized itinerant electrons but lack overall magnetization. Altermagnets have a unique spin-split band structure that arises from mechanisms other than relativistic effects like spin-orbit coupling.  Additionally, they do not exhibit invariance under the combined operations of parity and time-reversal. These properties enable the intrinsic generation of spin currents, which are crucial for advancing spintronic technologies.
The spin-splitting observed in altermagnets can reach substantial magnitudes, up to the scale of an electron volt \cite{Yang}.
Due to their intriguing physical characteristics and potential applications in spintronics, altermagnets have garnered significant attention within the condensed matter community.   The duality of real-space antiferromagnetism and reciprocal-space anisotropic spin polarization, akin to ferromagnetism, allows altermagnetic materials to exhibit many novel physical effects. These include giant magnetoresistance (GMR) effect \cite{Smejkal1}, piezomagnetic effect \cite{Ma}, spin-splitting torque \cite{Gonzalez-Hernandez,Bose,Bai,Karube}, tunneling magnetoresistance (TMR) effect \cite{Smejkal1,Shao}, time-reversal odd anomalous effect \cite{Smejkal2,Feng,Betancourt,Hou,Zhou1,Zhou2}, higher-order topological states \cite{Li1}, altermagnetic ferroelectricity \cite{Guo1}, nontrivial superconductivity \cite{Zhu,Li2,Ghorashi,Mondal,Fukaya,Pal},  quantum anomalous Hall effect \cite{Guo2}, electrical switching \cite{Chen1} and  strong spin-orbit coupling effects in light element altermagnetic materials \cite{Qu}. Additionally, predicted altermagnetic materials span metals, semimetals, and insulators, ensuring the potential realization of these novel physical effects in experiments \cite{Maznichenko,Gao1,Xiao1,Zeng,Liu}.

Thus far, a diverse array of materials has been categorized as altermagnets, including $\mathrm{RuO_2}$ \cite{Fedchenko}, $\mathrm{MnTe}$ \cite{ Krempasky,Osumi}, $\mathrm{MnF_2}$ \cite{Yuan} and others. However, the majority of documented altermagnets are three-dimensional (3D), with only limited reports on two-dimensional (2D) variants. Additionally, theoretical predictions have identified some 2D altermagnets, such as $\mathrm{V_2Te_2O}$ \cite{Cui}
and $\mathrm{RuF_4}$ \cite{Sødequist}. Studies have demonstrated that monolayer $\mathrm{MnP(S,Se)_3}$ can transition from antiferromagnetism to altermagnetism through the application of an electric field or by adopting a Janus structure, which disrupts inversion symmetry between sublattices \cite{Mazin2}.

The generation and control of spin currents lie at the heart of spintronics, a field that aims to exploit both the spin and charge attributes of carriers to control the properties of materials and devices
\cite{Wolf,Zutic,Bader,Schliemann}. 
Traditionally, spin current generation has relied on spin injection or pumping from neighboring ferromagnets \cite{ Datta,Gardelis,Schmidt,Hu,Tombros,Xiao}, optical injection \cite{Ganichev,Stevens} methods that rely on optical selection rules. Another significant approach uses heavy metals that exhibit strong spin-orbit coupling (SOC).  Here, it is worth noting that Rashba spin-orbit coupling (RSOC) \cite{Rashba1,Bychkov} (a specific form of SOC arising from the absence of surface inversion symmetry in electron confinement within quantum wells or heterostructures) is of notable significance in the advancing field of spintronics, enabling the development of novel devices with the ability to adjust the RSOC strength through external gate voltages or alternative approaches \cite{Nitta,Engels}. While these approaches have significantly advanced the field, they face inherent limitations. Ferromagnets introduce problematic stray magnetic fields that hinder device scaling and integration density, while conventional heavy metals often demand high power consumption and are restricted by the need for strong SOC, thereby limiting material design flexibility. These challenges underscore a critical need for alternative materials and mechanisms for robust and efficient spin current generation.

As discussed earlier, altermagnets possess a collinear antiferromagnetic-like ordering with zero net magnetization, yet they exhibit momentum-dependent spin splitting in their electronic band structure, even in the absence of SOC.  These properties make altermagnets promising platforms for efficient intrinsic spin current generation. Moreover, the absence of stray magnetic fields and the potential for ultrafast spin dynamics make altermagnets highly suited for dense integration in next-generation spintronic memory and logic devices.  Thus, understanding the spin transport properties in altermagnet systems is fundamentally important and strategically valuable for advancing scalable, energy-efficient technologies.

Inspired by the preceding discussion, we consider a 2D system that integrates Rashba spin-orbit coupling with altermagnetism \cite{Smejkal,Amundsen,Belashchenko,Chen2}, given their significant individual contributions to spin current generation.  In this model, the Rashba term breaks inversion symmetry, whereas the altermagnet term breaks  both time-reversal and combined parity–time (PT) symmetry. The discrete symmetries of the Hamiltonian significantly influence the behavior of spin currents. It is well established that in the Boltzmann transport, the presence of inversion and time reversal symmetry leads to the elimination of even and odd-order contributions of the electric field to the spin current, respectively \cite{Hamamoto}. As symmetries play a crucial role in determining the fate of spin currents,  we focus on how these broken symmetries in our system contribute to the generation of different order spin currents.

  With the symmetry analysis one can justify the existence of a finite spin current in noncentrosymmetric systems (absence of inversion symmetry) under thermodynamic equilibrium conditions (i.e. in absence of
an electric field), referred to as background equilibrium spin current \cite{Rashba2,Rashba3}. This spin current, while present, cannot transport  or accumulate electron spins.  The spin-orbit interaction acts as the driving force for spins, resulting in a pure persistent spin current. A significant amount of research has been dedicated to the study of persistent spin current \cite{Loss,Splettstoesser,Schutz,Usaj,Dolcini,Rossi}. Among various possibilities, the linear spin Hall current  has emerged as a prominent technique for generating and manipulating spin currents \cite{Kato,Day}.
Moreover, there has been a growing interest in producing nonlinear spin currents \cite{Yu,Hamamoto,Pan,Kapri,Wang}.
Nonlinear spin current can emerge in various scenarios, such as in a 2D crystal with Fermi surface anisotropy \cite{Yu} or in a noncentrosymmetric spin-orbit coupled system \cite{Hamamoto,Pan}. Our investigation therefore aims to explore the equilibrium, linear, and nonlinear spin currents within Boltzmann transport framework, particularly emphasizing how the interplay between Rashba and altermagnet parameters influences spin current behavior as a function of Fermi energy.

This paper is structured as follows: In Sec. (\ref{sec2}A) and Sec. (\ref{sec2}B), we discuss the formalism of spin current for a generic two-band system. Section  (\ref{sec2}C) encompasses the fundamental details regarding the 2D altermagnet Rashba system. Section (\ref{sec3})   includes associated findings on different order spin currents.  Lastly, in Sec. (\ref{sec4}), we offer concluding remarks and summarize our main results.

\section{Theory}
\label{sec2}
This section outlines a general formalism for spin currents of different orders in a two-band system subjected to an external electric field. We begin by exploring how the modified Fermi-Dirac distribution function under an external electric field allows us to define various orders of spin currents. Subsequently, we offer an overview of a 2D Rashba-coupled altermagnet system.

\subsection{Approximation in the carrier distribution function}
For the correction to the particle distribution function, our study adopts the approach of Ref. \cite{Pan}, where the derivation considers the local change in the equilibrium distribution function caused by the applied field. 

When a system is subjected to a  spatially uniform electric field ${\bf E}$,
the electron energy is changed to 
$\epsilon^\prime({\bf r}, {\bf k}) = \epsilon({\bf k}) + e {\bf E \cdot r}$, 
with ${\bf r}$ being the spatial coordinate.  Since the change in electron energy is very weak as compared to
the Fermi energy $\epsilon_F$,
the Fermi-Dirac distribution function 
$f({\bf r},{\bf k}) = 
(1 + e^{\beta [\epsilon^\prime({\bf r}, {\bf k})-\epsilon_F]} )^{-1}$
with $\beta = 1/(k_BT)$ can be expanded in a series of terms proportional 
to powers of  ${\bf E}$. Here, it is to be noted that the linear term is constrained
to replicate the solution of the Boltzmann transport equation (BTE). Hence,
the spatial coordinate must be written in the form  
${\bf r}= {\bf v}_{b} \tau$ with $\tau$ being 
the  relaxation time and ${\bf v}_{b}$ being the band velocity.  Thus, from the above discussion, the validity of our approximation requires
$e\tau |{\bf E}\cdot{\bf v}_{b}|/\epsilon_F \ll 1$, which in turn reduces to the condition $e\tau E v_b/\epsilon_F \ll 1$. With this consideration, the modified distribution function can be expressed as
\cite{Pan}
\begin{equation}
	\label{eqdfn}
	f(\epsilon, {\bf E}, \tau) = 
	\sum_{n}f_{n} = \sum_n \frac{[e\tau {\bf E} \cdot {\bf v}_{b}]^n}{n!} 
	\frac{\partial^n}{\partial \epsilon^n} [f_0(\epsilon)],
\end{equation}
where $n=0,1,2...$ and  $f_{0}(\epsilon) = 
(1 + e^{\beta [\epsilon({\bf k})-\epsilon_F]})^{-1}$ being the equilibrium 
distribution function in absence of the electric field. 
The term $ f_{n}(\sim E_{\eta}^n$, $\eta$: direction of electric field) denotes the $n$-th order deviation 
from the equilibrium distribution function $f_{0}(\epsilon)$ due to the external electric field.

\subsection{Redefinition of spin current}
In the presence of an external electric field 
${\bf E}$, a system with non-zero Berry curvature 
($\boldsymbol\Omega$) causes charge carriers to acquire an additional velocity transverse to the direction of the electric field. This additional velocity is commonly known as anomalous velocity.  
The velocity operator, including the anomalous term—$\hat{v}
_{i} = \hat{v}_{b,i} + \hat{v}
_{a,i}$, where $\hat{v}
_{b,i} = \frac{1}{\hbar} \frac{\partial H(\mathbf{k})}{\partial k_i}$ is the band velocity operator  ($H$: Hamiltonian, ${\bf k}$: wavevector), and $v_{a,i} = -(e/\hbar) \epsilon_{ijk} E_j \Omega_k$ ($\Omega_k$: Berry curvature in $k$ direction) represents the anomalous velocity in $i$ direction—is well established and widely employed in the literature \cite{Chang,Sundaram,Sodemann}, with the anomalous component being central to the anomalous Hall effect (AHE).
In the conventional definition of the spin current operator, only the band velocity contribution is considered, while the contribution from the anomalous velocity is completely neglected. The conventional definition of the spin current operator is given by:
$\hat{{v}}_{b,ij} = (\hat{v}_{b,i} \sigma_{j} + 
\sigma_{j} \hat{v}_{b,i} )/2$ ,
where the first index 
$i$ denotes the direction of propagation, the second index 
$j$ specifies the spin orientation of the charge carrier and
$\boldsymbol\sigma$ represents the Puauli's Matrices.
With inclusion of anomalous velocity,  the  redefined  
spin current operator is given by \cite{Kapri}
\begin{equation}  
	\hat{{v}}_{ij} = \hat{{v}}_{b,ij} + \hat{v}_{a,ij}, 
\end{equation} 
where $\hat{v}_{a,ij} = v_{a,i} \sigma_j$. With the above defined spin current operator, the total spin current  can be expressed as
\begin{equation}
	\mathcal{J}_{ij} = \frac{\hbar}{2 }  
	\sum_{n,s}
	\int \frac{ d^D{\bf k} }{(2\pi)^D} 
	\langle s, {\bf k}|\hat{{v}}_{ij}| s, {\bf k}\rangle f_{n}, 
\end{equation}
where $D$ denotes the spatial dimension of the system under consideration and  $s=\pm$ denotes the band indices. By separating the band and anomalous velocity contributions, the spin current of order 
$n$ arising from the band velocity can be expressed as
\begin{equation}
	\label{eqjb}
	\mathcal{J}_{b,ij}^{(n)} = \frac{\hbar}{2 } 
	\sum_{s} \int \frac{d^D{\bf k}}{(2\pi)^D} 
	\langle s, {\bf k}|\hat{{v}}_{b,ij}| s, {\bf k}\rangle f_{n}.
\end{equation}
It is important to note that for for $n=0$, the zeroth-order spin current ($\mathcal{J}^{(0)}_{xy}=\mathcal{J}^{(0)}_{b,ij}$), also known as the background spin current, originates exclusively from the band velocity, while the lowest-order contribution from the anomalous velocity appears at first order, since the anomalous velocity   itself depends linearly on electric field: ${\bf v}_a$  $\propto {\bf E}$ (note that the band velocity is independent of
electric field).
Consequently, the $(n+1)$-th order spin current originating from the anomalous velocity can be expressed as
\begin{equation}
	\label{eqja}
	\mathcal{J}_{a,ij}^{(n+1)} = \frac{\hbar}{2 } \sum_{s} 
	\int \frac{d^D{\bf k}}{(2\pi)^D} 
	\langle s, {\bf k}|\hat{{v}}_{a,ij}| s, {\bf k}\rangle f_{n}.
\end{equation}
It should be noted that the direction of the electric field in the spin current enters through the term 
$f_n$ (see Eq.~(\ref{eqdfn})) or $\hat{v}_{a,ij} = v_{a,i} \sigma_j$ 
(see the definition of the anomalous velocity discussed previously). To make this explicit, we introduce the index 
$\eta$ denoting the electric field direction in the spin current notation: 
$\mathcal{J}_{b, ij}^{(n),\eta}$ (for band velocity induced) and 
$\mathcal{J}_{a, ij}^{(n+1),\eta}$ (for anomalous velocity induced), which are used in presenting our numerical results in Secs.~(\ref{secB}) and (\ref{secC}). Finally, the total spin current of 
order $n$ can be calculated as  $\mathcal{J}_{ij}^{(n)}=\mathcal{J}_{b,ij}^{(n), \eta}+\mathcal{J}_{a,ij}^{(n),\eta}$, where 
$i$ denotes the propagation direction, 
$j$ the spin orientation, and 
$\eta$ the electric field direction.


\subsection{ Rashba coupled altermagnet system}
\begin{figure}
	\begin{center}	\includegraphics[width=90mm,height=90mm]{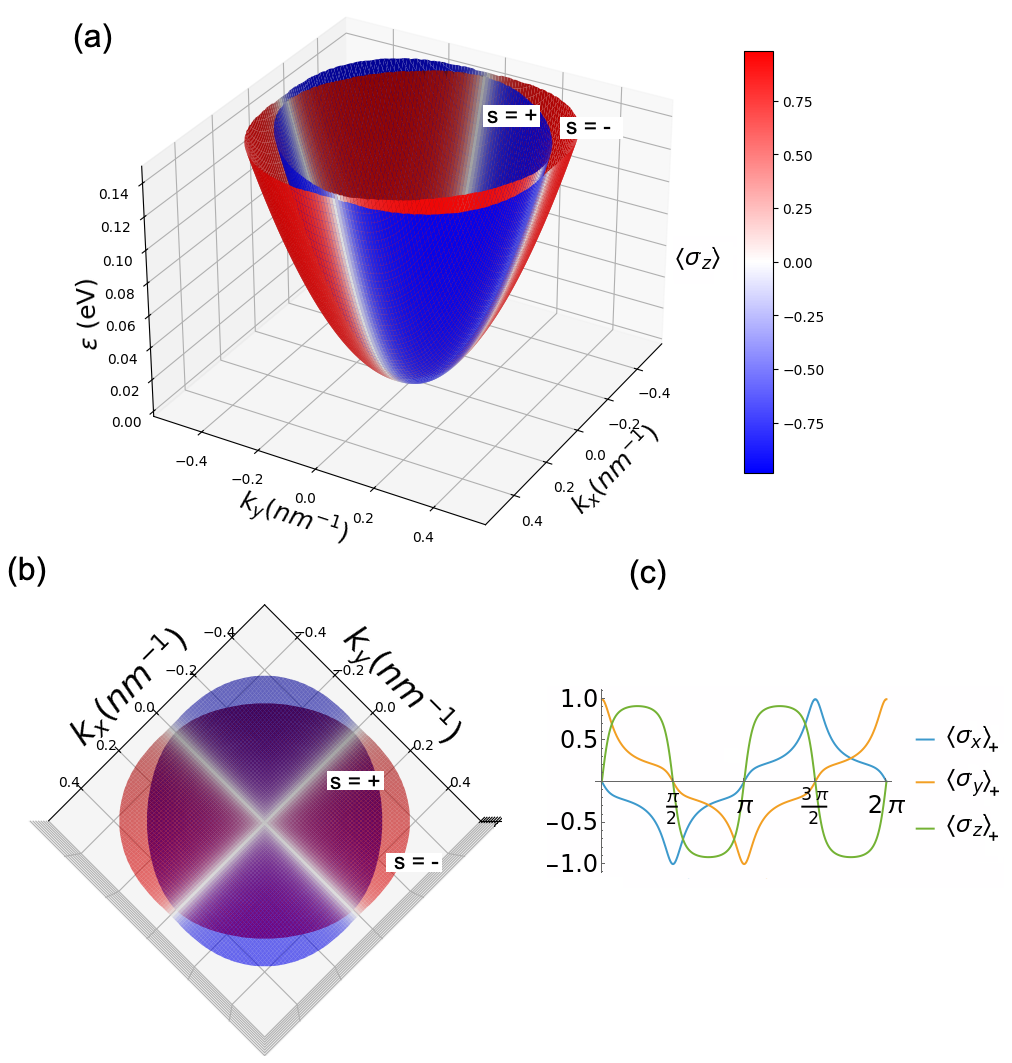}
		\caption {  (a) Band structure corresponding to the Hamiltonian in Eq. (\ref{eq1}) with parameters  $\mathrm{t_0 = \hbar^2 / (2m^*) = 0.761}$ $\mathrm{eV.nm^2}$, (with an effective mass of $\mathrm{m^* = 0.05m_e}$, where $\mathrm{m_e}$ is the free electron mass),  $\mathrm{t_j = 0.2t_0}$ and $\lambda = 0.02$ $\mathrm{eV.nm}$. The colormap indicates the out-of-plane spin polarization, ${\langle \sigma_z \rangle}_s$.  Panel (b) shows a top-down (constant energy) view of the full band dispersion, while panel (c) plots all components of the spin vector $\langle \boldsymbol{\sigma} \rangle_+$ as a function of the azimuthal angle $\phi$.  The anisotropic spin splitting reaches minima along azimuthal angles $\phi = n\pi/2$, where the spin vectors lie entirely within the $(k_x, k_y)$ plane. In contrast, maxima appear at $\phi = (2n + 1)\pi/4$, where the out-of-plane component $\langle\sigma_z\rangle$ is most pronounced.}
\label{Fig1}
	\end{center}
\end{figure}

The Hamiltonian for a two dimemnsional Rashba coupled altermagnet system is given by \cite{Smejkal,Chen2}
\begin{equation}
	\label{eq1}
	H=t_0\sigma_0k^2+2t_j\sigma_zk_xk_y+\lambda(\sigma_yk_x-\sigma_xk_y),
\end{equation}
where ${\bf k} = \{k \cos\phi, k \sin \phi\}$ is the electron’s wave vector,  $t_0$   is kinetic nearest neighbor hopping, $t_j$ is parameter that characterizes the altermagnetism strength , $\lambda$ is the strength of Rashba spin-orbit coupling, ${\bf \boldsymbol\sigma}=(\sigma_x,\sigma_y,\sigma_z)$ represents the usual Pauli's matrices and $\sigma_0$ is the $2\times2$ identity matrix. The two level energy spectrum of the Hamilonian in Eq. (\ref{eq1}) is obtained as 
\begin{equation}
	\epsilon_s({\bf k})=t_0k^2+s\Delta_{\bf k},
\end{equation}
with $\Delta_{ k}=k\sqrt{\lambda^2+t_j^2k^2\sin^22\phi}$. Both $\lambda$ and $t_j$ parameters contribute to the spin splitting ($2\Delta_{\bf k}$) between the two spin states $s=+$ and $s=-$, while $t_j$ is responsible for the anistropic spin splitting. The corresponding normalized eigen spinors for the spin-split dispersion are
\begin{eqnarray}
	\label{eqwf}
	|+, {\bf k}\rangle  = 
	\left[\begin{array}{c}
		\cos\frac{\theta_{\bf k}}{2}   \\
		i\sin\frac{\theta_{\bf k}}{2}e^{i\phi}
	\end{array} \right]\hspace{0.02in};\hspace{0.02in}
	|-,{\bf k}\rangle  = 
	\left[\begin{array}{c}
		\sin\frac{\theta_{\bf k}}{2}  \\
		-i\cos\frac{\theta_{\bf k}}{2}e^{i\phi}
	\end{array} \right],
\end{eqnarray}
where $\theta_{\bf k}=\cos^{-1}\big[\frac{t_jk^2\sin2\phi}{\Delta_{\bf k}}\big]$.
The spin orientation of an
electron with wave vector ${\bf k }$ in the Rashba coupled altermagnet system is
$\langle\boldsymbol\sigma\rangle_s=\{{\langle\sigma_x\rangle}_{s},{\langle\sigma_y\rangle}_s,{\langle\sigma_z\rangle}_s\} = s\{- \sin\theta_{\bf k}\sin \phi,  \sin\theta_{\bf k}\cos \phi, \cos \theta_{\bf k} \}$. Thus, the spin and linear
momentum lock in a way such that $\langle\boldsymbol\sigma\rangle_s \cdot {\bf k} = 0$.   
 It should be noted that the out-of-plane spin orientation 
($\langle\sigma_z\rangle_s$) vanishes when 
the altermagnet parameter 
$t_j$ is set to zero, while the in-plane spin orientations 
($\langle\sigma_x\rangle_s$ and ($\langle\sigma_y\rangle_s$) vanish when 
$\lambda$ is set to zero. Thus, the anisotropic out-of-plane spin polarization is a direct manifestation of altermagnetism, whereas the Rashba term is responsible for the in-plane polarization. 
Figure (\ref{Fig1}) depicts the band structure and spin texture of Rashba-coupled altermagnet systems, where the colormap represents the magnitude of the out-of-plane spin polarization, ${\langle \sigma_z \rangle}_s$. Unlike the Rashba spin-splitting,  the spin splitting here is anisotropic (please see Fig . (\ref{Fig1}a) and (\ref{Fig1}b)).  The spin splitting reaches its minimum along the azimuthal directions $\phi = n\pi/2$ (where $n$ is an integer), and reaches its maximum at $\phi = (2n + 1)\pi/4$. At the angles of minimum splitting, both spin vectors lie entirely within the $(k_x, k_y)$ plane (the white regime in the colormap, where  ${\langle \sigma_z \rangle}_s$ vanishes). In contrast, at the angles of maximum splitting, the spin vectors exhibit their strongest out-of-plane ($\langle\sigma_z\rangle$) component (please see Fig . (\ref{Fig1}c)).

Considering an excitation with energy $\epsilon$,
the wave
vector in the Rashba coupled altermagnet system for band $s$ can be written as
\begin{equation}k_s(\phi)=\Big[\frac{\lambda^2+2t_0\epsilon+s\big[\lambda^4+4t_0\lambda^2\epsilon+(2t_j\epsilon\sin2\phi)^2\big]^{\frac{1}{2}}}{2(t_0^2-t_j^2\sin^22\phi)}\Big]^{\frac{1}{2}},
\end{equation}
and the density of states  is given by
\begin{equation}
	\rho_s(\epsilon)=	\frac{1}{4\pi^2t_0}\int_{0}^{2\pi}\frac{1}{\Big|2+\frac{s[\lambda^2+2t_j^2	k_s(\phi)^2\sin^22\phi]}{t_0\big[\lambda^2	k_s(\phi)^2+t_j^2	k_s(\phi)^4\sin^22\phi\big]^{1/2}}\Big|}d\phi. 
\end{equation}
Moreover, the  Berry curvature
corresponding to $s$ band can be obtained as 
\begin{equation}
	\label{eq4}
	\Omega_{s}({\bf k})=s\frac{\lambda^2t_jk^2\sin2\phi}{2\Delta_{\bf k}^3}\hat{z},
\end{equation}
which  exhibits  anisotropic even-parity characteristic (a $d$-wave signature). 
The gap closing at the
origin results in  a singularity in $\Omega_{s}({\bf k})$.   In contrast, a counterpart
ferromagnet-Rashba model, $t_0\sigma_0k^2+\lambda(\sigma_yk_x-\sigma_xk_y
)+M\sigma_z$ ($M$ is the mass gap generated by breaking the
 time-reversal symmetry),
yields an isotropic Berry curvature near the $\Gamma$ point,
 consistent with the fundamentally isotropic 
$s$-wave character of ferromagnetism \cite{Nagaosa}. It is important to note that the Berry curvature vanishes when either  
$t_j=0$ or 
$\lambda=0$. This indicates that a nonzero Berry curvature arises only from the combined breaking of inversion and time reversal symmetries. 
Now, the generalized  velocity components for Rashba coupled altermagnet system arising from band velocity are given by
\begin{eqnarray*}
\label{EqA1}
\langle\hat{v}_{b,x}\rangle&=&\frac{1}{\hbar}[2t_0 k\cos\phi+2st_jk\cos\theta_{\bf k}\sin\phi+\lambda s\sin\theta_{\bf k}\cos\phi],\\\nonumber
	\langle\hat{v}_{b,y}\rangle&=&\frac{1}{\hbar}[2t_0 k\sin\phi+2st_jk\cos\theta_{\bf k}\cos\phi+\lambda s\sin\theta_{\bf k}\sin\phi]\nonumber,
\end{eqnarray*} while the velocity components arising from anomalous velocity are obtained as
\begin{eqnarray*}
\langle\hat{v}_{a,x}\rangle=-\frac{e}{\hbar} E_y\Omega_z,\hspace{1mm}
\langle\hat{v}_{a,y}\rangle=\frac{e}{\hbar} E_x\Omega_z\nonumber.
\end{eqnarray*}
Further, the spin velocity components resulting from the band velocity are derived as follows
\begin{eqnarray*}	\label{EqA2}\langle\hat{v}_{b,xx}\rangle&=-&\frac{1}{\hbar}st_0k\sin\theta_{\bf k}\sin2\phi,\\\nonumber
	\langle\hat{v}_{b,xy}\rangle&=&\frac{1}{\hbar}[\lambda+2st_0k\sin\theta_{\bf k}\cos^2\phi],\\\nonumber
		\langle\hat{v}_{b,xz}\rangle&=&\frac{2}{\hbar}[t_j k\sin\phi+st_0k\cos\theta_{\bf k}\cos\phi],\\\nonumber
		\langle\hat{v}_{b,yx}\rangle&=&-\frac{1}{\hbar}[\lambda+2st_0k\sin\theta_{\bf k}\sin^2\phi], \\\nonumber
\langle\hat{v}_{b,yy}\rangle&=&\frac{1}{\hbar}st_0k\sin\theta_{\bf k}\sin2\phi\\\nonumber	\langle\hat{v}_{b,yz}\rangle&=&\frac{2}{\hbar}[t_j k\cos\phi+st_0k\cos\theta_{\bf k}\sin\phi].\\\nonumber
\end{eqnarray*}
It is important to note that the spin velocity  resulting from band component with out-of-plane polarization ($\langle\hat{v}_{b,xz}\rangle$ and $\langle\hat{v}_{b,yz}\rangle$) vanishes in the pure Rashba case, since for 
 $t_j=0$, $\cos\theta_k=0$. We find that these  components ($\langle\hat{v}_{b,xz}\rangle$ and $\langle\hat{v}_{b,yz}\rangle$) are responsible for the spin-Hall current (a transverse  spin current with out-of-plane polarization) in altermagnetic system, which we discuss in detail later.
Now the spin velocity components resulting from the
anomalous velocity are obtained as
\begin{eqnarray*}
\label{EqA3}	\langle\hat{v}_{a,xx}\rangle&=&s\frac{e}{\hbar}E_y\Omega_z\sin\theta_k\sin\phi,\\\nonumber	\langle\hat{v}_{a,xy}\rangle&=&-s\frac{e}{\hbar}E_y\Omega_z\sin\theta_k\cos\phi, \\\nonumber	
\langle\hat{v}_{a,xz}\rangle&=&-s\frac{e}{\hbar}E_y\Omega_z\cos\theta_k,\\\nonumber
 \langle\hat{v}_{a,yx}\rangle&=&-s\frac{e}{\hbar}E_x\Omega_z\sin\theta_k\sin\phi,\\\nonumber
\langle\hat{v}_{a,yy}\rangle&=&s\frac{e}{\hbar}E_x\Omega_z\sin\theta_k\cos\phi, \\\nonumber
\langle\hat{v}_{a,yz}\rangle&=&s\frac{e}{\hbar}E_x\Omega_z\cos\theta_k.
\end{eqnarray*}
which exist only in systems with nonzero Berry curvature. Here as well, the spin velocity component with out-of-plane polarization contributes to the spin-Hall current; however, this contribution is negligible compared to that from the band component (elaborated later). 



\section{Results and Discussions}
\label{sec3}

This section presents the results for different order spin 
currents along with the associated discussions. The plots for spin current
are presented as a function of Fermi energy ($\epsilon_F$) for different values of $t_j$ and $\lambda$. For all plots, we consider $\mathrm{t_0=\hbar^2/(2m^*)=0.761}$ $\mathrm{eV.nm^2}$  with $\mathrm {m^*=0.05 m_e}$ being the effective mass ($\mathrm{m_e}$: electronic mass). Further, we consider three values of altermagnet and Rashba parameter as  $\mathrm{t_j=[0.2, 0.4, 0.6]t_0}$ and $\lambda=0.02, 0.04, 0.06$ $\mathrm {eV.nm}$, respectively.  The temperature for all
the plots is fixed at $T = 0$  {\rm K}.

\subsection{Background spin current}
\label{sec5}

First we present the results for the nonpropagating background spin current \cite{Rashba2}. Using  Eq. (\ref{eqjb}) with $n=0$, we find that  $\mathcal{J}^{(0)}_{xx}=\mathcal{J}^{(0)}_{yy}=0$. Similarly, we find $\mathcal{J}^{(0)}_{xz}=\mathcal{J}^{(0)}_{yz}=0$.  On the other hand, we
obtain $\mathcal{J}^{(0)}_{xy}\neq0$ and $\mathcal{J}^{(0)}_{yx}\neq0$. At zero temperature $\mathcal{J}^{(0)}_{xy}$ and $\mathcal{J}^{(0)}_{yx}$
have the following forms (for details, please see Eq. (\ref{EqA1}) in Appendix \ref{appA})
\begin{eqnarray}
\label{eqbc}
		\mathcal{J}^{(0)}_{xy}
		&=&	\frac{\lambda}{16\pi^2}\sum_s\int_0^{2\pi}\Big[[k_s^{F}(\phi)]^2 +2st_0\Bigg(\frac{\Delta_s^F(\phi)}{t_j^2\sin^2(2\phi)}\\\nonumber&-&\lambda^2\frac{\tanh ^{-1}\Big[\frac{t_j[k_s^{F}(\phi)]^2\sin(2\phi)}{\Delta_s^F(\phi)}\Big]}{[t_j\sin(2\phi)]^{3}}\Bigg)\cos^2\phi  \Big]d\phi ,\\\nonumber
  \mathcal{J}^{(0)}_{yx}
		&=&	-\frac{\lambda}{16\pi^2}\sum_s\int_0^{2\pi}\Big[[k_s^{F}(\phi)]^2 + 2st_0\Bigg(\frac{\Delta_s^F(\phi)}{t_j^2\sin^2(2\phi)}\\\nonumber&-&\lambda^2\frac{\tanh ^{-1}\Big[\frac{t_j[k_s^{F}(\phi)]^2\sin(2\phi)}{\Delta_s^F(\phi)}\Big]}{[t_j\sin(2\phi)]^{3}}\Bigg)\sin^2\phi  \Big] d\phi,\end{eqnarray}
  where $\Delta_{ s}^F(\phi)=k_s^F(\phi)[\lambda^2+t_j^2[k_s^F(\phi)]^2\sin^2(2\phi)]^{1/2}$ with $k_s^F(\phi)$ being the wavevector at $\epsilon=\epsilon_F$ for band $s$. Evaluating the $\phi$ integration in Eq. (\ref{eqbc}), we obtain that $\mathcal{J}^{(0)}_{xy}=-\mathcal{J}^{(0)}_{yx}$. \begin{figure}\includegraphics[width=88mm,height=100mm]{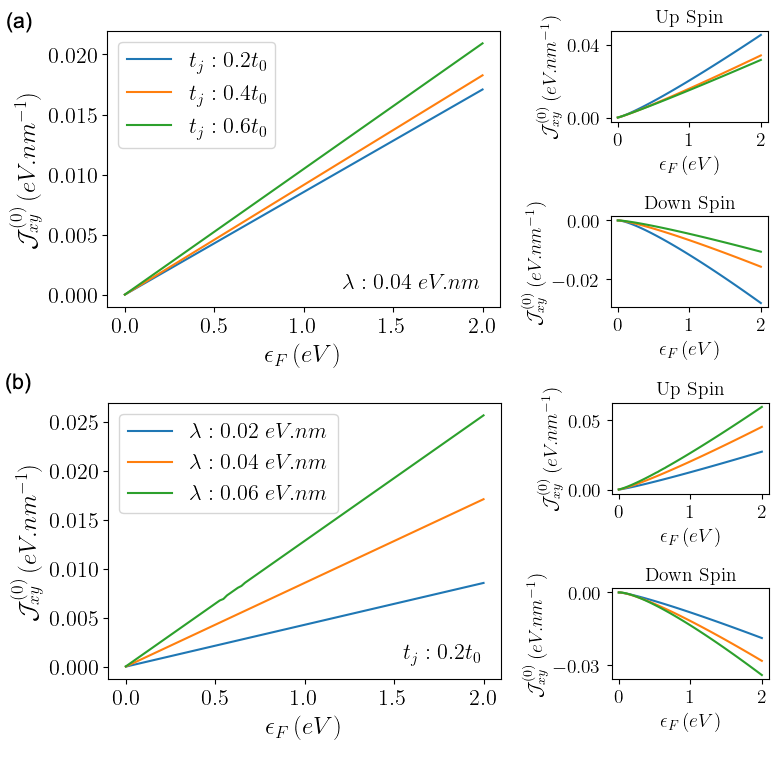}
\caption {The background spin current $\mathcal{J}^{(0)}_{xy}$ (in $\mathrm{eV.nm^{-1}}$) as a function of
 Fermi energy $\epsilon_F$ (in $\mathrm{eV}$)  (a) for different values of $t_j$ with $\lambda=0.04 $ $\mathrm{eV.nm}$ and (b) for different values of $\lambda$ with $t_j=0.2 $ $t_0$. The corresponding spin-split current is shown in the right panel.  It should be noted that
 in the pure Rashba case ($t_j=0$, $\lambda \neq 0$, $M=0$) and in the gapped Rashba case ($t_j=0$, $\lambda \neq 0$, $M\neq0$), the background spin current becomes independent of the Fermi energy for $\epsilon_F > 0$ and $\epsilon_F > M$, respectively, and is given by $\mathcal{J}^{(0)}_{xy}=(\lambda^3{m^*}^2)/(6\pi\hbar^4)$. For $\lambda = 0.02, 0.04,$ and $0.06$ $\mathrm{eV.nm}$, $\mathcal{J}^{(0)}_{xy}=(\lambda^3{m^*}^2)/(6\pi\hbar^4)$ takes the values $0.19 \times 10^{-6}$, $1.52 \times 10^{-6}$, and $5.13 \times 10^{-6}$ $\mathrm{eV. nm^{-1}}$, respectively—significantly weaker than in the Rashba–altermagnet system. } 
\label{Fig2}
\end{figure}
  Here, it is to be noted that in  the pure Rashba case ($t_j=0$, $\lambda \neq 0$, $M=0$) as well as in  the gapped Rashba case ($t_j=0$, $\lambda \neq 0$, $M \neq 0$), one can find $\mathcal{J}^{(0)}_{xx}=\mathcal{J}^{(0)}_{yy}=\mathcal{J}^{(0)}_{xz}=\mathcal{J}^{(0)}_{yz}=0$ and $\mathcal{J}^{(0)}_{xy}=-\mathcal{J}^{(0)}_{yx}\neq0$ \cite{Rashba2,Kapri}. 
  In the case of altermagnet Rashba ($t_j\neq 0$, $\lambda\neq 0$), one might expect the  $t_j$
  term to contribute to 
$\mathcal{J}^{(0)}_{xz}$
  or 
$\mathcal{J}^{(0)}_{yz}$, unlike the pure Rashba case ($t_j=0$). However, this contribution is absent because the terms involving 
$t_j$
  that would affect the spin current
$\mathcal{J}^{(0)}_{xz}$
  and 
$\mathcal{J}^{(0)}_{yz}$
  vanish upon integration over 
 $\phi$  (see the expressions for $\langle\hat{v}_{b,xz}\rangle_s$ and $\langle\hat{v}_{b,yz}\rangle_s$).  Though the altermagnet term influences the behavior of the background spin current, it does not create it on its own, as $\mathcal{J}^{(0)}_{xy}=\mathcal{J}^{(0)}_{yx}=0$ for $\lambda=0$. Thus, the background spin current is a characteristic of the inversion-breaking system, as it vanishes when inversion symmetry is restored.
  
  In Fig. (\ref{Fig2}), we show the plots for background spin current  $\mathcal{J}^{(0)}_{xy}$  as a function of
 Fermi energy $\epsilon_F$. The upper plot represents the spin current for different values of $t_j$ with $\lambda=0.04 $ $\mathrm{eV.nm}$, while the lower panel represents the spin current for different values of $\lambda$ with $t_j=0.2 $ $t_0$.
The background spin current for Rashba coupled altermagnet systems appears nearly linear as a function of Fermi energy. The right panel displays the individual spin contributions to the background spin current, clearly illustrating how the total spin current varies with Fermi energy. It is worth noting that in the pure Rashba case ($t_j=0$, $\lambda \neq 0$, $M=0$) and in the gapped Rashba case ($t_j=0$, $\lambda \neq 0$, $M\neq0$), the background spin current becomes independent of the Fermi energy for $\epsilon_F > 0$ and $\epsilon_F > M$, respectively,  and is given by $\mathcal{J}^{(0)}_{xy}=(\lambda^3{m^*}^2)/(6\pi\hbar^4)$ \cite{Rashba2,Kapri}. 
Additionally, we observe that both 
$t_j$
  and $\lambda$ contribute to an enhancement of the background spin current.
  
  \subsection{Linear Spin current}
  \label{secB}
  Here we present the results of the linear spin current for all possible combinations. First, we consider the linear spin current arising from the band velocity (calculated from Eq. (\ref{eqjb}) with $n=1$), where we find $\mathcal{J}_{b,xx}^{(1),x}=\mathcal{J}_{b,xx}^{(1),y}=\mathcal{J}_{b,xy}^{(1),x}=\mathcal{J}_{b,xy}^{(1),y}=\mathcal{J}_{b,xz}^{(1),x}=\mathcal{J}_{b,yx}^{(1),x}=\mathcal{J}_{b,yx}^{(1),y}=\mathcal{J}_{b,yy}^{(1),x}=\mathcal{J}_{b,yy}^{(1),y}=\mathcal{J}_{b,yz}^{(1),y}=0$ and $\mathcal{J}_{b,xz}^{(1),y}\neq0$ and $\mathcal{J}_{b,yz}^{(1),x}\neq0$. Therefore, linear currents arising from band velocity vanish if the propagation direction aligns with either the polarization direction or the direction of the electric field and hence the spin currents always have out-plane spin polarization. These linear spin currents transverse to the electric field  with out-of-plane spin polarization ($\mathcal{J}_{b,yz}^{(1),x}$ and $\mathcal{J}_{b,xz}^{(1),y}$) can be interpreted as spin Hall currents. In the pure 2D Rashba case ($t_j=0$, $\lambda\neq 0$, $M=0$) \cite{Rashba2}, as well as in the gapped 2D Rashba case ($t_j=0$, $\lambda\neq 0$, $M\neq0$)\cite{Kapri}, there are no linear currents originating from band velocity.
  In our system, 
$t_j$
is responsible for the band-induced linear spin Hall current (explained in detail later) and hence, the altermagnetic system can be considered as a promising candidate for the generation of spin Hall current.

   At zero temperature, the band induced linear spin Hall currents have the following forms (for details, please see Eq. (\ref{EqA2}) in Appendix \ref{appA})
 \begin{eqnarray}
	\mathcal{J}^{(1),y}_{b,xz}
	&=&\frac{-e\tau E_y}{4\hbar\pi^2}\sum_s\int_0^{2\pi}\Big[2t_0k\sin\phi+2st_j\\\nonumber&\times&k\cos\theta_{k}\cos\phi+\lambda s\sin\theta_{k}\sin\phi\Big]_{\epsilon=\epsilon_F(\phi)}\\\nonumber&\times&\Big[st_0k\cos\theta_{k}\cos\phi+t_jk\sin\phi\Big]_{\epsilon=\epsilon_F(\phi)}\\\nonumber&\times& |k\frac{\partial k} {\partial \epsilon}|_{\epsilon=\epsilon_F(\phi)}  d\phi,\\\nonumber
 \mathcal{J}^{(1),x}_{b,yz}
	&=&\frac{-e\tau E_x}{4\hbar\pi^2}\sum_s\int_0^{2\pi}\Big[2t_0k\cos\phi+2st_j\\\nonumber&\times&k\cos\theta_{k}\sin\phi+\lambda s\sin\theta_{k}\cos\phi\Big]_{\epsilon=\epsilon_F(\phi)} \\\nonumber&\times&\Big[st_0k\cos\theta_{k}\sin\phi+t_jk\cos\phi\Big]_{\epsilon=\epsilon_F(\phi)} \\\nonumber&\times&|k\frac{\partial k} {\partial \epsilon}|_{\epsilon=\epsilon_F(\phi)} d\phi,
\end{eqnarray}
where we find that $\mathcal{J}^{(1),y}_{b,xz}=\mathcal{J}^{(1),x}_{b,yz}$ in case of $E_x=E_y$.

\begin{figure}
\includegraphics[width=89mm,height=100mm]{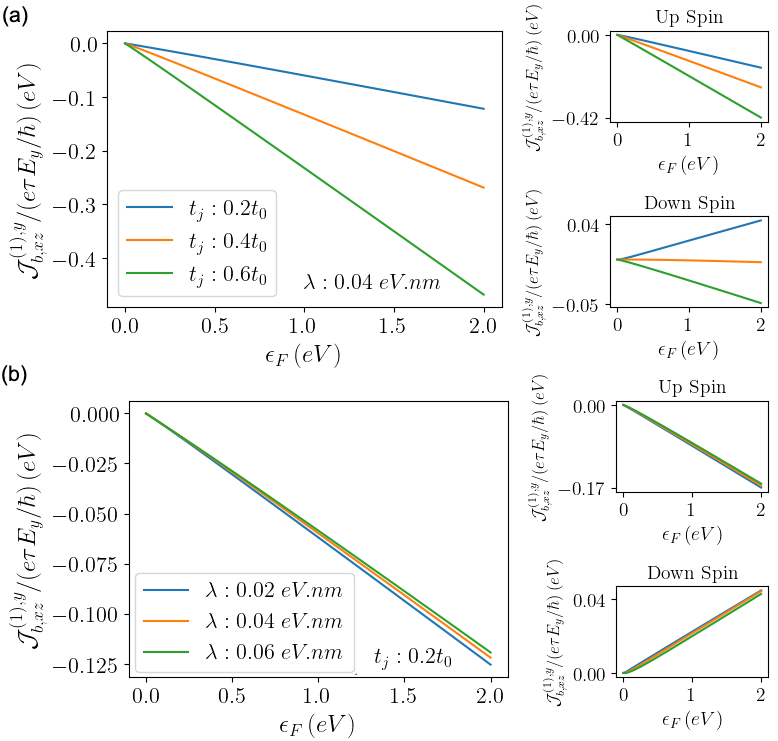}
\caption {The plots for $\mathcal{J}^{(1),y}_{b,xz}/(e\tau E_y/\hbar)$ (in $\mathrm{eV}$), ($\mathcal{J}^{(1),y}_{b,xz}$: linear spin Hall current  from  band velocity)
as a function of
 Fermi energy $\epsilon_F$ (in $\mathrm{eV}$)  (a) for different values of $t_j$ with $\lambda=0.04 $ $\mathrm{eV.nm}$ and (b) for different values of $\lambda$ with $t_j=0.2 $ $t_0$. The right panel displays the corresponding spin-split current.  It should be noted that, in both the pure and gapped Rashba cases, the band-velocity contribution to the linear spin Hall current vanishes. }
\label{Fig3}
\end{figure}

 In Fig. (\ref{Fig3}), we present the plots for  $\mathcal{J}^{(1),y}_{b,xz}/(e\tau E_y/\hbar)$ as a function of Fermi energy.  
 The upper plot shows the spin current for different values of $t_j$ with $\lambda=0.04 $ $\mathrm{eV.nm}$, while the lower plot displays the spin current for different values of 
$\lambda$ with $t_j=0.2 $ $t_0$.
 The linear spin current originating from the band velocity increases in magnitude with the Fermi energy, displaying a linear dependence. To elucidate the contributions of different spin channels to the spin current, the spin-split current is presented in the right panel. As expected, the altermagnetic parameter 
($t_j$)
  leads to an enhancement of the first-order spin current magnitude. Conversely, the Rashba  parameter 
($\lambda$) induces a slight suppression of the current, although its overall contribution remains comparatively weak. This is because the contributions from the spin-up and spin-down channels nearly cancel each other as 
$\lambda$ varies.

Now, we consider the linear spin current arising from the anomalous velocity (calculated from Eq. (\ref{eqja}) with $n=0$), where we find that $\mathcal{J}_{a,xx}^{(1),x}=\mathcal{J}_{a,xx}^{(1),y}=\mathcal{J}_{a,xy}^{(1),x}=\mathcal{J}_{a,xy}^{(1),y}=\mathcal{J}_{a,xz}^{(1),x}=\mathcal{J}_{a,yx}^{(1),x}=\mathcal{J}_{a,yx}^{(1),y}=\mathcal{J}_{a,yy}^{(1),x}=\mathcal{J}_{a,yy}^{(1),y}=\mathcal{J}_{a,yz}^{(1),y}=0$ and $\mathcal{J}_{a,xz}^{(1),y}\neq0$ and $\mathcal{J}_{a,yz}^{(1),x}\neq0$. Similar to the linear spin currents arising from band velocity, the linear spin currents in this case also consistently exhibit out-of-plane spin polarization. These spin currents can also be interpreted as spin Hall currents arising from a nonzero Berry curvature, which requires both the altermagnetic parameter and the Rashba coupling to be nonzero.
At zero temperature, the  linear spin Hall  currents arising from anomalous velocity have the
following forms (for details, please see Eq. (\ref{EqA3}) in Appendix \ref{appA})
\begin{eqnarray}
	\mathcal{J}^{(1),y}_{a,xz}
		=
		\frac{eE_y\lambda^2}{32\pi^2}\sum_s
	\int_0^{2\pi}&\Big(&\frac{1}{\lambda^2+[t_jk_s^F(\phi)\sin(2\phi)]^2}\\\nonumber&-&\frac{1}{\lambda^2}\Big)   d\phi \\\nonumber	
 \mathcal{J}^{(1),x}_{a,yz}
		=
		-\frac{eE_x\lambda^2}{32\pi^2}\sum_s
	\int_0^{2\pi}&\Big(&\frac{1}{\lambda^2+[t_jk_s^F(\phi)\sin(2\phi)]^2}\\\nonumber&-&\frac{1}{\lambda^2}\Big)   d\phi
\end{eqnarray}
and hence we obtain $\mathcal{J}^{(1),y}_{a,xz}=-\mathcal{J}^{(1),x}_{a,yz}$ with the consideration $E_x=E_y$.  Since we obtain $\mathcal{J}^{(1),y}_{b,xz}=\mathcal{J}^{(1),x}_{b,yz}$ and $\mathcal{J}^{(1),y}_{a,xz}=-\mathcal{J}^{(1),x}_{a,yz}$, the total spin current (considering both the band and anomalous components) $\mathcal{J}^{(1),y}_{xz}$ ($\mathcal{J}^{(1),y}_{b,xz}$+ $\mathcal{J}^{(1),y}_{a,xz}$) and $\mathcal{J}^{(1),x}_{yz}$ ($\mathcal{J}^{(1),x}_{b,yz}$+ $\mathcal{J}^{(1),x}_{a,yz}$) have different values (even with the consideration $E_x=E_y$).  
\begin{figure}
\includegraphics[width=88mm,height=105mm]{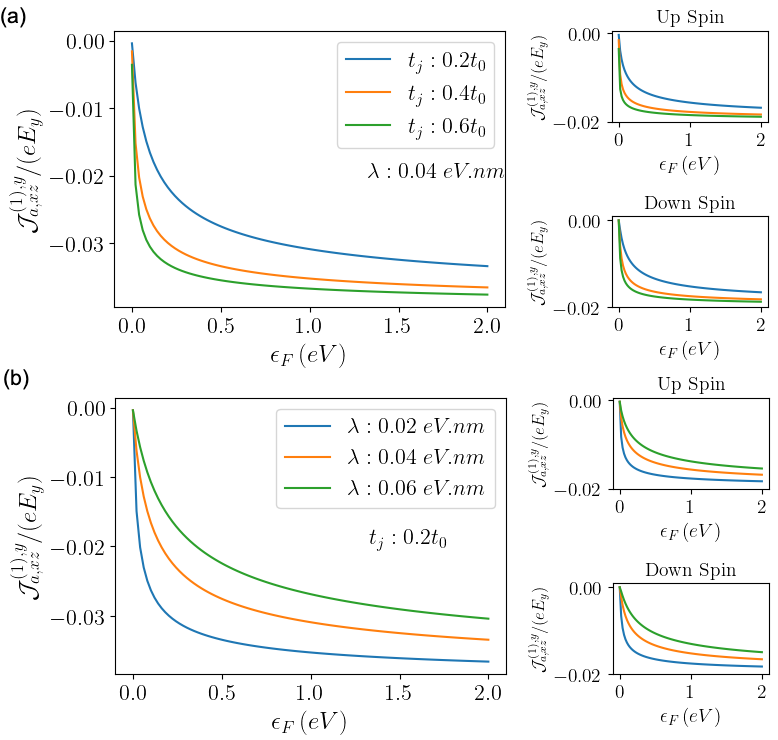}
\caption {The plots for $\mathcal{J}^{(1),y}_{a,xz}/(eE_y)$ (dimensionless) ($\mathcal{J}^{(1),y}_{a,xz}$: linear spin Hall current  from anomalous velocity) as a function of
 Fermi energy $\epsilon_F$ (in $\mathrm{eV}$)  (a) for different values of $t_j$ with $\lambda=0.04 $ $\mathrm{eV.nm}$ and (b) for different values of $\lambda$ with $t_j=0.2 $ $t_0$. The corresponding spin-split current is shown in the right panel.  It should be noted that, in the pure Rashba case, $\mathcal{J}^{(1),y}_{a,xz}/(eE_y)$ is independent of the Fermi energy for $\epsilon_F>0$ and takes a constant value: $-(1/8\pi) \approx -0.0398$. For  gapped Rashba case with $\epsilon_F \gg M$, $\mathcal{J}^{(1),y}_{a,xz}/(eE_y)$ approaches the same constant value: $-(1/8\pi)$.
 } 
\label{Fig4}
\end{figure}

In Fig. (\ref{Fig4}), we present the plot for the linear spin Hall current $\mathcal{J}^{(1),y}_{a,xz}/(eE_y)$ as a function of Fermi energy $\epsilon_F$. As earlier, the upper plot displays the spin current for different values of $t_j$ with $\lambda=0.04 $ $\mathrm{eV.nm}$, while the lower plot shows the spin current for different values of 
$\lambda$ with $t_j=0.2 $ $t_0$.
The magnitude of the linear spin Hall current arising from anomalous velocity increases with Fermi energy, though its behavior differs from that of the spin Hall current induced by band velocity. At higher Fermi energies, the anomalous-velocity-induced spin Hall current tends to saturate, in contrast to the band-induced spin Hall current, which continues to grow. 
As before, 
$t_j$
 enhances the magnitude of the linear spin Hall current induced by anomalous velocity; however, increasing the Rashba strength leads to the opposite effect.

 We now provide a brief discussion regarding the band-induced linear spin Hall current, which is absent in pure or gapped Rashba systems. As mentioned earlier, that in the Boltzmann transport the odd-order spin currents can only arise when time-reversal symmetry is broken \cite{Hamamoto}. In our Rashba-coupled altermagnet system, the altermagnet parameter breaks time-reversal symmetry and thereby enables the linear spin Hall current from both the band and anomalous velocity.
 Within Boltzmann transport for the Rashba model, introducing a Zeeman-like term ($M\sigma_z$)
breaks time-reversal symmetry and generates a spin Hall current through a nonzero Berry curvature—i.e., via the anomalous velocity. Notably, once the spin Hall current is evaluated in the presence of this time-reversal–breaking term, setting $\epsilon_F>>M $ (or $M \rightarrow 0$) recovers the well-known Kubo formalism (which inherently incorporates Berry curvature effects) result: $\mathcal{J}^{(1),y}_{a,xz}/(eE_y)=-(1/8\pi)$ for $M=0$ \cite{Sinova,Kapri}. Therefore, in Rashba systems, the spin Hall effect originates solely from the anomalous velocity rather than the band velocity. 
In altermagnets, however, the altermagnet parameter not only breaks time-reversal symmetry and induces a nonzero Berry curvature, but also gives rise to additional band-induced spin velocity components—
($\langle\hat{v}_{b,xz}\rangle$ and $\langle\hat{v}_{b,yz}\rangle$) (as discussed earlier)—which are absent in the pure Rashba system.
These spin velocity components, generated by the time-reversal symmetry-breaking parameter 
$t_j$, lead to a band-induced spin Hall current—a distinctive characteristic of altermagnets.

\begin{figure}
\includegraphics[width=80mm,height=60mm]{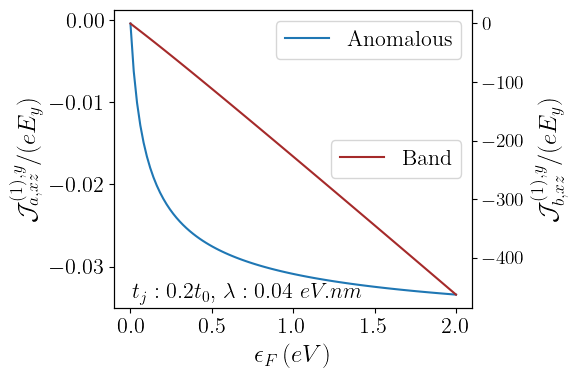}
\caption {The plots for $\mathcal{J}^{(1),y}_{b,xz}/(eE_y)$ (dimensionless) and $\mathcal{J}^{(1),y}_{a,xz}/(eE_y)$ (dimensionless) as a function of
Fermi energy $\epsilon_F$ (in eV) with $\tau=2.5$ ps. This confirms that
the band velocity plays the dominant role in realizing spin
Hall currents in Rashba-coupled altermagnets.  The above statement remains valid even under significant variations in material parameters.}
\label{Fig5}
\end{figure}

Next, we compare the linear spin Hall currents induced by band velocity and anomalous velocity  with $\tau=2.5$ $\mathrm{ps}$ in Fig. (\ref{Fig5}), which shows 
$\mathcal{J}^{(1),y}_{b,xz}/(eE_y)$ is significantly larger than 
$\mathcal{J}^{(1),y}_{a,xz}/(eE_y)$. 
To facilitate the comparison between 
$\mathcal{J}^{(1),y}_{b,xz}/(eE_y)$ and $\mathcal{J}^{(1),y}_{a,xz}/eE_y$) in Fig. (\ref{Fig5}), the result shown in Fig. (\ref{Fig3}) has been multiplied by the factor $\tau/\hbar= 0.1502\times 10^{16}\times\tau$ $\mathrm {(e V. s)^{-1}}$, which for 
$\tau=2.5$ $\mathrm{ps}$ gives a multiplicative factor $0.3755 \times 10^4$ $\mathrm {eV}^{-1}$.
It is important to note that the relaxation time in altermagnetic materials is not a fixed constant; rather, it varies depending on the specific material, temperature, and dominant relaxation mechanisms. Accurate values of $\tau$ must therefore be obtained through theoretical modeling and experimental measurements. Recent studies report a broad range of relaxation times in altermagnets, from approximately 
$0.05$ $\mathrm{ps}$ to
$50$ $\mathrm{ps}$ \cite{Sun}, corresponding to $\tau/\hbar$
 values in the range of 
$0.0075\times 10^4-7.5\times 10^4$ $\mathrm {e V^{-1}}$.
Within this relevant range of multiplicative factor ($\tau/\hbar$), our key conclusion remains robust: the spin Hall current contribution from the band velocity term 
($\mathcal{J}^{(1),y}_{b,xz}/(eE_y)$), is significantly larger than that from the anomalous velocity term, 
$\mathcal{J}^{(1),y}_{a,xz}/(eE_y)$.
Further, we verified that even with significantly varied material parameters—such as very large 
$\lambda$ (up to 
$0.5$ $\mathrm{eV.nm}$; for comparison, 
$\mathrm{In_xGa_{1-x}As/GaAs}$ quantum dots exhibit 
$\lambda$ in the range 
$0.08-0.12$ $\mathrm{eV.nm}$ \cite{Huang}, while the 
$\mathrm{Bi/Ag(111)}$ surface alloy shows a large Rashba strength of 
$0.305$ $\mathrm{eV.nm}$ \cite{Gierz}) and 
$t_j$ (up to $0.9 t_0$)—the conclusion that the band contribution dominates over the anomalous contribution in the linear spin Hall current remains valid. This indicates that, although the Berry curvature contributes to the spin Hall current, the band velocity plays the dominant role in realizing spin Hall currents in Rashba-coupled altermagnets. 

 It is important to emphasize that Ref. \cite{Chen2} examined the spin Hall current in the same Rashba-coupled altermagnet system within the Kubo formalism. Since this framework inherently incorporates Berry curvature effects, it naturally attributes the spin Hall response to the anomalous velocity contribution. In contrast, our semiclassical Boltzmann approach reveals that the band velocity alone can generate a spin Hall current, independent of Berry curvature. This distinction highlights the central finding of our work (band velocity induced spin Hall current), which cannot be captured within the Kubo formalism. 

\subsection{Nonlinear spin current}
\label{secC}
In the previous subsection, we showed that the transverse linear spin current is finite, while the longitudinal component vanishes. The leading nonzero contribution to the longitudinal spin current is quadratic in the electric field 
${\bf E}$, a hallmark of systems with broken inversion symmetry. In this subsection, we analyze the quadratic spin current under different configurations of spin orientation, charge transport direction, and applied electric field. We begin by presenting cases where the quadratic spin current arises purely from the band velocity. At zero temperature, the expression for the non-linear current arising from the band velocity takes the following form (for details, please see Eq. (\ref{EqA5}) in Appendix \ref{appA})
\begin{eqnarray}
\mathcal{J}_{b,ij}^{(2),\eta}&=&\frac{\hbar e^2\tau^2E_{\eta}^2}{16\pi^2}\sum_s\int_0^{2\pi}\Big[\frac{\partial}{\partial \epsilon}[\mathcal{H}_s(\epsilon)]\Big]_{\epsilon=\epsilon_F(\phi)}d\phi,
\end{eqnarray}
with $\mathcal{H}_s(\epsilon)=|k\frac{\partial k}{\partial \epsilon}|\langle\hat{v}_{b,ij}\rangle_s\langle\hat{v}_{b,\eta}\rangle_s^2$.\\
{\bf Results for $\mathcal{J}_{b,xx}^{(2),\eta}$, $\mathcal{J}_{b,yy}^{(2),\eta}$ }:
Now we shall present
results for quadratic spin currents, $\mathcal{J}_{b,xx}^{(2),\eta}$ and $\mathcal{J}_{b,yy}^{(2),\eta}$
when both propagation and polarization are in the same
direction. Here we find $\mathcal{J}_{b,xx}^{(2),x}=\mathcal{J}_{b,xx}^{(2),y}=\mathcal{J}_{b,yy}^{(2),x}=\mathcal{J}_{b,yy}^{(2),y}=0$. Thus, the band velocity does not contribute to the nonlinear spin currents when the propagation and polarization directions are the same. The same phenomenon is also observed in both pure and gapped Rashba systems \cite{Kapri}.\\
{\bf Results for $\mathcal{J}_{b,xy}^{(2),\eta}$, $\mathcal{J}_{b,yx}^{(2),\eta}$ }: Here we present the results for quadratic spin currents  $\mathcal{J}_{b,xy}^{(2),\eta}$, $\mathcal{J}_{b,yx}^{(2),\eta}$, where we consider the in-plane polarization, but the propagation and polarization direction are perpendicular to each other. Interestingly, we find $\mathcal{J}_{b,xy}^{(2),x}=-\mathcal{J}_{b,yx}^{(2),y}\neq0$ and 
$\mathcal{J}_{b,xy}^{(2),y}=-\mathcal{J}_{b,yx}^{(2),x}\neq0$. Therefore, the band velocity contributes to both the longitudinal and transverse non linear spin currents.
\begin{figure}
\includegraphics[width=87mm,height=100mm]{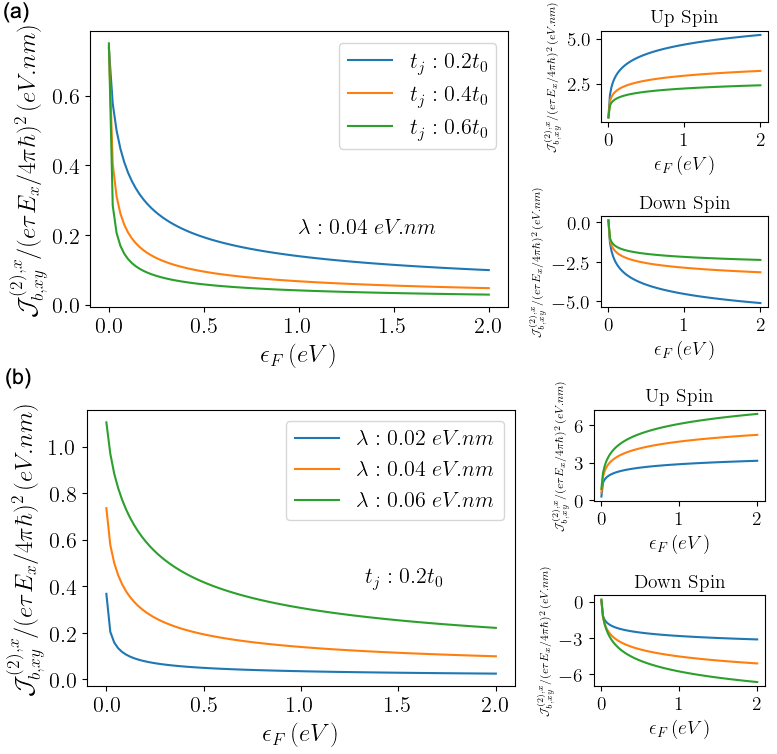}
\caption {The plots for $\mathcal{J}^{(2),x}_{b,xy}/(e\tau E_x/4\pi\hbar)^2$ (in $\mathrm{eV.nm}$) ($\mathcal{J}^{(2),x}_{b,xy}$: nonlinear longitudinal spin current  from band velocity) as a function of Fermi energy $\epsilon_F$ (in $\mathrm{eV}$)  (a) for different values of $t_j$ with $\lambda=0.04 $ $\mathrm{eV.nm}$ and (b) for different values of $\lambda$ with $t_j=0.2 $ $t_0$. The right panel displays the corresponding spin-split current. It should be noted that, in the pure Rashba case, $\mathcal{J}^{(2),x}_{b,xy}/(e\tau E_x/4\pi\hbar)^2$ is independent of the Fermi energy and assumes a constant value of $-2\pi\lambda$ for $\epsilon_F>0$. For $\lambda = 0.02, 0.04,$ and $0.06$ $\mathrm{eV.nm}$, the corresponding values are $-0.126$, $-0.251$, and $-0.377$ $\mathrm{eV.nm}$, respectively.} 
\label{Fig6}
\end{figure}
\begin{figure}
\includegraphics[width=88mm,height=100mm]{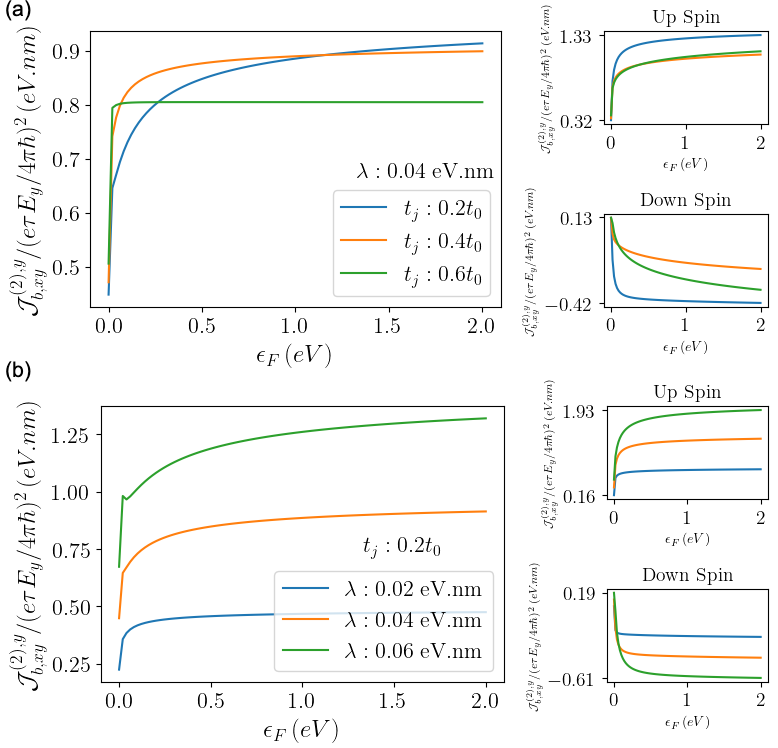}
\caption {The plots for $\mathcal{J}^{(2),y}_{b,xy}/(e\tau E_y/4\pi\hbar)^2$ (in $\mathrm{eV.nm}$) ($\mathcal{J}^{(2),y}_{b,xy}$: nonlinear transverse spin current  from band velocity) as a function of Fermi energy $\epsilon_F$ (in $\mathrm{eV}$)  (a) for different values of $t_j$ with $\lambda=0.04 $ $\mathrm{eV.nm}$ and (b) for different values of $\lambda$ with $t_j=0.2 $ $t_0$. The corresponding spin-split current is shown in the right panel. It should be noted that, in the pure Rashba case, $\mathcal{J}^{(2),y}_{b,xy}/(e\tau E_y/4\pi\hbar)^2$ is independent of the Fermi energy for $\epsilon_F>0$ and assumes a constant value of $2\pi\lambda$. For $\lambda = 0.02, 0.04,$ and $0.06$ $\mathrm{eV.nm}$, the corresponding values are $0.126$, $0.251$, and $0.377$ $\mathrm{eV.nm}$, respectively.} 
\label{Fig7}
\end{figure}
In Figs. (\ref{Fig6}) and (\ref{Fig7}), we present the plots  for $\mathcal{J}^{(2),x}_{b,xy}/(e\tau E_x/4\pi\hbar)^2$ and $\mathcal{J}^{(2),y}_{b,xy}/(e\tau E_y/4\pi\hbar)^2$ as a function of Fermi energy $\epsilon_F$. As earlier cases, the upper plots show the spin current for different values of $t_j$ with $\lambda=0.04 $ $\mathrm{eV.nm}$, while the lower plots display the spin current for different values of 
$\lambda$ with $t_j=0.2 $ $t_0$. To clarify the individual contributions of distinct spin channels to the overall spin current, the corresponding spin-split current is depicted in the right panel. For longitudinal current, the nonlinear spin current originating from band velocity reaches its maximum at lower Fermi energy values, then decreases as the Fermi energy increases, eventually saturating at higher Fermi energies.  Furthermore, the altermagnetic term reduces the longitudinal nonlinear current, while the Rashba term enhances it. However, the transverse nonlinear spin current originating from band velocity exhibits a different behavior as a function of Fermi energy. It shows an increasing trend  with Fermi energy and eventually saturates at higher Fermi energies. Furthermore, the nonlinear transverse spin current from band velocity exhibits a nonmonotonic dependence on the altermagnet parameter, whereas the Rashba spin-orbit coupling consistently enhances the transverse spin current.  
 Notably, in our previous study on the gapped Rashba system, the nonlinear spin current from band velocity also exhibited saturation at higher Fermi energies with sharp peaks at the gap edges \cite{Kapri}. Moreover, we obtain that for a pure Rashba system, both the longitudinal and transverse nonlinear spin currents arising from band velocity are Fermi-energy independent for $\epsilon_F>0$, and are given by $\mathcal{J}^{(2),x}_{b,xy}/(e\tau E_x/4\pi\hbar)^2=-\mathcal{J}^{(2),y}_{b,xy}/(e\tau E_y/4\pi\hbar)^2=-2\pi\lambda$. It is worth noting that, unlike in the Rashba case, for the Rashba–altermagnet system $\mathcal{J}^{(2),x}_{b,xy}/(e\tau E_x/4\pi\hbar)^2 \neq -\mathcal{J}^{(2),y}_{b,xy}/(e\tau E_y/4\pi\hbar)^2$. 
 \\
{\bf Results for $\mathcal{J}_{b,xz}^{(2),\eta}, \mathcal{J}_{b,yz}^{(2),\eta}$}:  Now, we consider the case where the polarization is out of the plane. Here, we obtain that $\mathcal{J}_{b,xz}^{(2),\eta}=\mathcal{J}_{b,yz}^{(2),\eta}=0$. It should be noted that although the $t_j$ term contributes to spin velocity operators $\langle\hat{v}_{b,xz}\rangle_s$ and $\langle\hat{v}_{b,yz}\rangle_s$ (unlike the pure 2D Rashba case), it does not contribute to $\mathcal{J}_{b,xz}^{(2),\eta}$  and $\mathcal{J}_{b,yz}^{(2),\eta}$ due to the $\phi$ integration. Thus, the spin currents with out-of-plane polarization due to band velocity are always of linear order in the electric field.

Now, we consider the nonlinear spin current arising from the anomalous velocity. At zero temperature, the expression for the nonlinear spin current due to anomalous velocity takes the following form (for details, please see Eq. (\ref{EqA6}) in Appendix \ref{appA})
\begin{eqnarray}
\mathcal{J}_{a,ij}^{(2),\eta}=-\frac{\hbar e\tau E_{\eta}}{8\pi^2}\sum_s \int_0^{2\pi}&\Big[&|k\frac{\partial k}{\partial \epsilon}|\langle\hat{v}_{a,ij}\rangle_s\\\nonumber&\times &\langle\hat{v}_{b,\eta}\rangle_s\Big]_{\epsilon=\epsilon_F(\phi)}d\phi.
\end{eqnarray}
\\
\begin{figure}
\includegraphics[width=88mm,height=105mm]{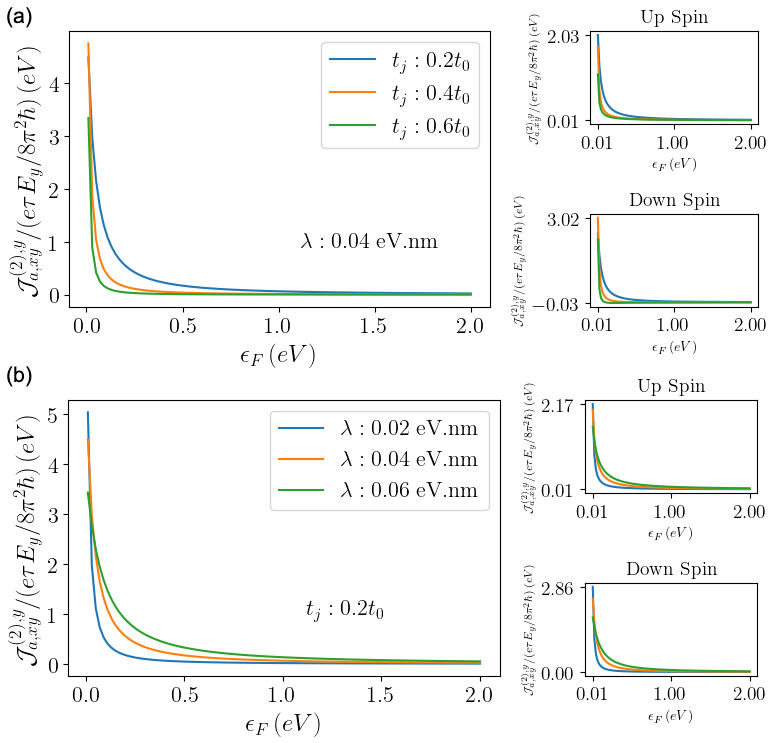}
\caption {The plots for  $\mathcal{J}^{(2),y}_{a,xy}/(e\tau E_y/8\pi^2\hbar)$ (in eV) ($\mathcal{J}^{(2),y}_{a,xy}$: nonlinear transverse spin current from anomalous velocity)  as a function of Fermi energy $\epsilon_F$ (in eV)  (a) for different values of $t_j$ with $\lambda=0.04 $ $\mathrm{eV.nm}$ and (b) for different values of $\lambda$ with $t_j=0.2 $ $t_0$. The right panel displays the corresponding spin-split current. It should be noted that, in the pure Rashba case, the nonlinear spin current from the anomalous velocity is zero.}
\label{Fig8}
\end{figure}
{\bf Results for $\mathcal{J}_{a,xx}^{(2),y}, \mathcal{J}_{a,xy}^{(2),y}, \mathcal{J}_{a,yx}^{(2),x}, \mathcal{J}_{a,yy}^{(2),x}$}:
Here we consider the nonlinear transverse spin current 
arising from anomalous velocity with in-plane polarization 
where we find that $\mathcal{J}_{a,xx}^{(2),y}=\mathcal{J}_{a,yy}^{(2),x}=0$ and $\mathcal{J}_{a,xy}^{(2),y}=\mathcal{J}_{a,yx}^{(2),x}\neq0$ with the consideration $E_x=E_y$. Here it should be noted that for gapped 2D Rashba case \cite{Kapri}, the scenario is opposite, where $\mathcal{J}_{a,xx}^{(2),y}=\mathcal{J}_{a,yy}^{(2),x}\neq0$ and $\mathcal{J}_{a,xy}^{(2),y}=\mathcal{J}_{a,yx}^{(2),x}=0$. For the 2D gapped Rashbha case, the mass gap $M$ (which is absent in our case) is responsible for $\mathcal{J}_{a,xx}^{(2),y}=\mathcal{J}_{a,yy}^{(2),x}\neq0$, whereas in our case, the altermagnet $t_j$ term is responsible for $\mathcal{J}_{a,xy}^{(2),y}=\mathcal{J}_{a,yx}^{(2),x}\neq0$ (which is absent in 2D gapped Rashba case).  Further, in the pure Rashba case, the anomalous velocity contributes no nonlinear spin current. In Fig. (\ref{Fig8}), we present the plot for nonlinear spin current $\mathcal{J}^{(2),y}_{a,xy}/(e\tau E_y/8\pi^2\hbar)$  as a function of Fermi energy $\epsilon_F$. The plots show that as 
$\epsilon_F$
  approaches $0$, the nonlinear spin current arising from anomalous velocity peaks, then begins to decrease, and eventually becomes zero at higher Fermi energies.  Furthermore, $t_j$
  reduces the nonlinear spin current originating from anomalous velocity, whereas the Rashba term enhances it. \\
{\bf Results for $\mathcal{J}_{a,xz}^{(2),y}, \mathcal{J}_{a,yz}^{(2),x}$}: Here we consider the nonlinear transverse spin current 
arising from anomalous velocity with out of plane polarization, 
where we find that $\mathcal{J}_{a,xz}^{(2),y}=\mathcal{J}_{a,yz}^{(2),x}=0$. Thus, the spin current with out-of-plane polarization is always linear with respect to the electric field. The nonlinear transverse spin currents consist of both band and anomalous components. 
 Next, we compare the nonlinear transverse spin currents induced by the band and anomalous velocities. For a direct comparison between $\mathcal{J}^{(2),y}_{b,xy}/(e\tau E_y/8\pi^2\hbar)$ and $\mathcal{J}^{(2),y}_{a,xy}/(e\tau E_y/8\pi^2\hbar)$, the result in Fig.~(\ref{Fig7}) should be multiplied by $(e\tau E_y)/(2\hbar)$, which for $\tau = 2.5$ $\mathrm{ps}$ and $E_y = 10^3$ $\mathrm{V/m}$ yields a factor of $0.19 \times 10^{-2}$ $\mathrm{nm^{-1}}$. The chosen values of $E_y$ and $\tau$ are consistent with the validity of our distribution function approximation.  With this consideration, the dominant contribution to the transverse nonlinear spin current—whether from band velocity or anomalous velocity—depends entirely on the Fermi-energy regime. Specifically, at lower Fermi energies the anomalous component dominates, whereas at higher Fermi energies the band contribution prevails, as the anomalous component tends to vanish.

\begin{table}
	\begin{center}
	\label{tab1}
		\begin{tabular}{ | c | c | c | c| c | c |} 
			\hline
			 Spin current & ${\bf E}=0$ & $\eta=x$ (B) & $\eta=x$ (A) & $\eta=y$ (B) & $\eta=y$ (A)\\
			\hline
			$\mathcal{J}_{xx}^{(0)} $ & 0 & NA & NA & NA & NA  \\
			\hline
			$\mathcal{J}_{xy}^{(0)} $ & Finite & NA & NA & NA & NA \\
			\hline
			$\mathcal{J}_{xz}^{(0)} $ & 0 & NA & NA& NA & NA	\\
			\hline
			$\mathcal{J}_{xx}^{1,\eta} $ & NA & 0 &	0& 0 & 0	\\
			\hline
			$\mathcal{J}_{xy}^{1,\eta} $ & NA & 0 &	0  & 0 & 0 \\
			\hline
			$\mathcal{J}_{xz}^{1,\eta} $ & NA & 0 & 0& Finite & Finite  \\
			\hline
			$\mathcal{J}_{xx}^{2,\eta} $ & NA & 0 & 0 & 0 & 0 \\
			\hline
			$\mathcal{J}_{xy}^{2,\eta} $ & NA & Finite & 0 & Finite & Finite	\\
			\hline
			$\mathcal{J}_{xz}^{2,\eta} $ & NA & 0 & 0& 0 & 0	\\
			\hline
		\end{tabular}\\
		
		*NA: Not Applicable, *B: Band component contribution, \\
  *A: Anomalous component contribution
		\label{tab1}
		\caption{Nature of spin currents in 2D  Rashba coupled altermagnet system for different orientations of electric field ${\bf E}$. }
	\end{center}
\end{table}
All the above results of spin current (whether it is zero
or non-zero)  are tabulated in Table I
 and  are fully consistent with symmetry analysis.  As mentioned earlier, inversion and time-reversal symmetries suppress even and odd order spin-current contributions, respectively. In our system, inversion symmetry is broken by the Rashba term, while time-reversal symmetry is broken by the altermagnetic parameter. Consequently, the Rashba term is responsible for nonzero background and nonlinear spin currents, whereas the altermagnet term originates the linear spin  current.
Further, we know that the Rashba term produces in-plane spin polarization (the in-plane spin orientations 
($\langle\sigma_x\rangle_s=\langle\sigma_y\rangle_s=0$ at 
$\lambda=0$ ), while the altermagnet parameter generates out-of-plane polarization ($\langle\sigma_z\rangle_s=0$ at $t_j=0$). Thus, even-order spin currents are finite only for in-plane polarization (no out-of-plane polarization for background and nonlinear spin currents), whereas odd-order spin currents are finite only for out-of-plane polarization (no in-plane polarization for linear spin current). Moreover, spin and momentum are locked such that $\langle \boldsymbol{\sigma} \rangle_s \cdot {\bf k} = 0$, ensuring that the spin polarization is always perpendicular to the propagation direction, and consequently, spin currents with coinciding propagation and polarization directions vanish. Furthermore, since time-reversal symmetry breaking is linked to transverse responses, the altermagnet parameter yields only transverse linear spin currents (i.e., spin Hall currents) and no longitudinal component. This explanation accounts for all the vanishing and non-vanishing components listed in Table 1.
\section{Conclusion}
\label{sec4}
We have investigated the background, linear, and nonlinear spin currents in two-dimensional Rashba spin–orbit coupled altermagnet systems, analyzing their distinct contributions to spin current generation within a semiclassical framework that incorporates Berry curvature–induced anomalous velocity. The Rashba term breaks inversion symmetry, while the altermagnet term breaks time-reversal  symmetry; together, these broken symmetries play a crucial role in determining the spin current behavior.   The study primarily focuses on the interplay between the altermagnet parameter ($t_j$) and the Rashba parameter ($\lambda$) in relation to spin currents as a function of Fermi energy ($\epsilon_F$).

We find that the background spin current emerges solely in the presence of spin-orbit coupling, with the altermagnet term ($t_j$) significantly influencing its behavior but unable to generate it independently. It consistently exhibits in-plane polarization, with the propagation direction perpendicular to this polarization. The background spin current increases nearly linearly with Fermi energy and is enhanced by both the altermagnet and Rashba parameters.

The most interesting finding of our study is the band velocity induced linear spin Hall current—absent in a simple Rashba-coupled system ($t_j=0$)—underscoring its potential as a promising source of spin Hall current generation. Linear spin currents are always transverse with out-of-plane polarization and can be interpreted as Spin Hall currents that arise from both the band and anomalous velocities, with band velocity playing the primary role in their realization in Rashba-coupled altermagnets. It is important to note that the anomalous component of the spin Hall current vanishes at $\lambda = 0$, whereas the band-induced spin Hall current remains nonzero. The magnitude of the band-driven linear spin Hall current increases almost linearly with Fermi energy, while the magnitude of anomalous velocity contribution initially rises but saturates at higher energies. Additionally, the altermagnet parameter enhances the absolute value of linear spin Hall current, whereas the Rashba parameter suppresses it.

The leading nonzero longitudinal spin current is quadratic in the electric field ${\bf E}$, a signature of broken inversion symmetry. Unlike linear spin currents, nonlinear spin currents feature both longitudinal and transverse components with in-plane polarization, where the propagation direction is perpendicular to the polarization. The longitudinal component from band velocity decreases and saturates with increasing $\epsilon_F$, while the transverse component grows before saturating. In contrast, the transverse contribution from anomalous velocity diminishes with $\epsilon_F$ and almost vanishes at high $\epsilon_F$. The altermagnet parameter $t_j$ suppresses the nonlinear longitudinal current, while the Rashba parameter $\lambda$ enhances it. The transverse current from band velocity is enhanced by $\lambda$ and shows a non-monotonic dependence on $t_j$, whereas $t_j$ reduces and $\lambda$ increases the transverse current from anomalous velocity.  
This comprehensive study of the interplay between Rashba and altermagnet parameters may provide valuable insights for controlling spin currents in Rashba spin-orbit coupled altermagnet systems.

\section{Acknowledgement}
Author acknowledges Prof. T. K. Ghosh for useful discussions.
\section{Data availability}
All data supporting the findings of this study are presented within the figures in the article.
\appendix
\begin{widetext}



\section{Detail calculation of spin currents}
\label{appA}
At zero temperature the background spin current $\mathcal{J}^{(0)}_{xy}$ can be calculated as follows
\begin{eqnarray}
\label{EqA1}		\mathcal{J}^{(0)}_{xy}&=&\frac{\hbar}{2}\frac{1}{(2\pi)^2}\sum_s\int d^2k \langle\hat{v}_{b,xy}\rangle_s \\\nonumber&=&
		\frac{1}{2}\frac{1}{(2\pi)^2}\sum_s\int\int [\lambda+2st_0k\sin\theta_k\cos^2\phi]k dk d\phi \\\nonumber
		&=&
		\frac{1}{2}\frac{1}{(2\pi)^2}\sum_s\Big[\lambda\int\Big(\int k dk\Big) d\phi+ 2st_0\int\Big(\int k^2\sin\theta_k dk\Big)\cos^2\phi  d\phi\Big] \\\nonumber
		&=&
		\frac{1}{2}\frac{1}{(2\pi)^2}\sum_s\Big[\lambda\int_0^{2\pi}\Big(\int_0^{k_s^{F}(\phi)} k dk\Big) d\phi+ 2s\lambda t_0\int_0^{2\pi}\Big(\int_0^{k_s^{F}(\phi)} \frac{k^2}{\sqrt{\lambda^2+t_j^2k^2\sin^2(2\phi)}}  dk\Big)\cos^2\phi  d\phi \Big]\\\nonumber
			&=&	\frac{\lambda}{16\pi^2}\sum_s\int_0^{2\pi}\Big[[k_s^{F}(\phi)]^2 +2st_0\Bigg(\frac{\Delta_s^F(\phi)}{t_j^2\sin^2(2\phi)}-\lambda^2\frac{\tanh ^{-1}\Big[\frac{t_j[k_s^{F}(\phi)]^2\sin(2\phi)}{\Delta_s^F(\phi)}\Big]}{[t_j\sin(2\phi)]^{3}}\Bigg)\cos^2\phi  \Big]d\phi. \\\nonumber						
				\end{eqnarray}

    At zero temperature the linear spin Hall current $\mathcal{J}^{(1),y}_{b,xz}$ arising from band velocity can be calculated as follows
\begin{eqnarray}
\label{EqA2}	\mathcal{J}^{(1),y}_{b,xz}&=&\frac{\hbar}{2}\frac{1}{(2\pi)^2}\sum_s\int d^2k \langle\hat{v}_{b,xz}\rangle f_1\\\nonumber
	&=&\frac{e\tau E_y\hbar}{8\pi^2}\sum_s\int d^2k \langle\hat{v}_{b,xz}\rangle \langle\hat{v}_{b,y}\rangle \frac{\partial f_0}{\partial \epsilon}\\\nonumber
	&=-&\frac{e\tau E_y}{4\pi^2\hbar}\sum_s\int\int |k \frac{\partial k}{\partial \epsilon }| [st_0k\cos\theta_k\cos\phi+t_jk\sin\phi][2t_0k\sin\phi+2st_jk\cos\theta_k\cos\phi+\lambda s\sin\theta_k\sin\phi] \delta(\epsilon-\epsilon_F)d\epsilon d\phi\\\nonumber
	&=&
	\frac{-e\tau E_y}{4\hbar\pi^2}\sum_s\int_0^{2\pi}\Big[[st_0k\cos\theta_{k}\cos\phi+t_jk\sin\phi][2t_0k\sin\phi+2st_jk\cos\theta_{k}\cos\phi+\lambda s\sin\theta_{k}\sin\phi]|k\frac{\partial k} {\partial \epsilon}|\Big]_{\epsilon=\epsilon_F(\phi)} d\phi.
\end{eqnarray}

    At zero temperature the linear spin Hall current $\mathcal{J}^{(1),y}_{a,xz}$ arising from anomalous velocity can be calculated as follows
\begin{eqnarray}
	\label{EqA3}
 \mathcal{J}^{(1),y}_{a,xz}&=&\frac{\hbar}{2}\frac{1}{(2\pi)^2}\sum_s\int d^2k \langle\hat{v}_{a,xz}\rangle_s \\\nonumber&=-&
	\frac{\hbar}{2}\frac{1}{(2\pi)^2}\sum_s\int\int [s\frac{e}{\hbar}E_y\Omega_z\cos\theta_k]k dk d\phi \\\nonumber
		&=-&
	\frac{eE_y\lambda^2 t_j^2}{4}\frac{1}{(2\pi)^2}\sum_s\int_0^{2\pi}\Big(\int_0^{k_s^{F}(\phi)} \frac{k}{(\lambda^2+t_j^2k^2\sin^2(2\phi))^2}  dk\Big)\sin(2\phi)^2 d\phi \\\nonumber	
	&=&	\frac{eE_y\lambda^2}{32\pi^2}\sum_s
	\int_0^{2\pi}\Big(\frac{1}{\lambda^2+[t_jk_s^F(\phi)\sin(2\phi)]^2}-\frac{1}{\lambda^2}\Big)   d\phi. \\\nonumber
	\end{eqnarray}
 At zero temperature the non linear spin  current  $\mathcal{J}_{b,ij}^{(2),\eta}$ arising from band velocity can be calculated as follows
 \begin{eqnarray}
 \label{EqA4}
\mathcal{J}_{b,ij}^{(2),\eta}&=&\frac{\hbar}{2}\frac{1}{(2\pi)^2}
	\sum_{s} \int d^2{\bf k}
	\langle \hat{{v}}_{b,ij} \rangle f_2 \\\nonumber
 &=&\frac{\hbar(e\tau E_{\eta})^2}{16\pi^2}
	\sum_{s} \int\int |k \frac{\partial k}{\partial \epsilon }| 
	\langle \hat{{v}}_{b,ij} \rangle_s \langle v_{b,\eta}\rangle_s^2
	\frac{\partial^2 f_0}{\partial \epsilon^2}d\epsilon d\phi
 \\\nonumber
 &=&\frac{\hbar(e\tau E_{\eta})^2}{16\pi^2}
	\sum_{s} \int\int \mathcal{H}_s(k(\epsilon)) 
	\frac{\partial^2 f_0}{\partial \epsilon^2}d\epsilon d\phi\\\nonumber
 \end{eqnarray}
 Performing integration by parts 
with $\mathcal{H}_s(k(\epsilon))=|k\frac{\partial k}{\partial \epsilon}|\langle\hat{v}_{b,ij}\rangle_s\langle\hat{v}_{b,\eta}\rangle_s^2$, we have
\begin{eqnarray}
\label{EqA5}
\mathcal{J}_{b,ij}^{(2),\eta}&=-&\frac{\hbar(e\tau E_{\eta})^2}{16\pi^2}
	\sum_{s} \int\int\Big[ \frac{\partial }{\partial \epsilon}\mathcal{H}_s(k(\epsilon)
	\frac{\partial f_0}{\partial \epsilon}\Big]d\epsilon d\phi
 \\\nonumber
 &=&\frac{\hbar(e\tau E_{\eta})^2}{16\pi^2}
	\sum_{s} \int \Big[\frac{\partial }{\partial \epsilon}[\mathcal{H}_s(k(\epsilon)] \Big]_{\epsilon=\epsilon_F(\phi)}d\phi.
\end{eqnarray}

 At zero temperature the non linear spin  current  $\mathcal{J}_{a,ij}^{(2),\eta}$ arising from anomalous velocity can be calculated as follows
 \begin{eqnarray}
 \label{EqA6}
\mathcal{J}_{a,ij}^{(2),\eta}&=&\frac{\hbar}{2}\frac{1}{(2\pi)^2}
	\sum_{s} \int d^2{\bf k}
	\langle \hat{{v}}_{a,ij} \rangle f_1 \\\nonumber
 &=&\frac{\hbar e\tau E_{\eta}}{8\pi^2}
	\sum_{s} \int\int\Big[ |k \frac{\partial k}{\partial \epsilon }| 
	\langle \hat{{v}}_{a,ij} \rangle_s \langle v_{b,\eta}\rangle_s
	\frac{\partial f_0}{\partial \epsilon}\Big]d\epsilon d\phi
 \\\nonumber
 &=-&\frac{\hbar e\tau E_{\eta}}{8\pi^2}
	\sum_{s} \int\Big[ |k \frac{\partial k}{\partial \epsilon }| 
	\langle \hat{{v}}_{a,ij} \rangle_s \langle v_{b,\eta}\rangle_s\Big]_{\epsilon=\epsilon_F(\phi)} d\phi.
 \end{eqnarray}
 
		\end{widetext}

\end{document}